\documentclass[aps,prl,twocolumn,superscriptaddress,floats]{revtex4}
\usepackage{txfonts}
\usepackage{amssymb}
\usepackage{graphicx}
\usepackage{epstopdf}
\usepackage{multirow}
\usepackage{float}
\usepackage{indentfirst}

\begin{document}
\title{Coexistence of ferromagnetic and stripe-type antiferromagnetic spin fluctuations in YFe$_2$Ge$_2$}
\author{Hongliang Wo}
\author{Qisi Wang}
\author{Yao Shen}
\author{Xiaowen Zhang}
\author{Yiqing Hao}
\author{Yu Feng}
\author{Shoudong Shen}
\author{Zheng He}
\author{Bingying Pan}
\affiliation{State Key Laboratory of Surface Physics and Department of Physics, Fudan University, Shanghai 200433, China}
\author{Wenbin Wang}
\affiliation{Institute of Nanoelectronic Devices and Quantum Computing, Fudan University, Shanghai 200433, China}
\author{K. Nakajima}
\author{S. Ohira-Kawamura}
\affiliation{Materials and Life Science Division, J-PARC Center, Tokai, Ibaraki 319-1195, Japan}
\author{P. Steffens}
\author{M. Boehm}
\affiliation{Institut Laue-Langevin, 71 Avenue des Martyrs, 38042 Grenoble Cedex 9, France}
\author{K. Schmalzl}
\affiliation{Forschungszentrum J$\ddot{u}$lich GmbH. J$\ddot{u}$lich Centre for Neutron Science at ILL,
71 Avenue des Martyrs, 38000 Grenoble, France}
\author{T. R. Forrest}
\affiliation{Diamond Light Source, Harwell Campus, Didcot OX11 0DE, United Kingdom}
\author{M. Matsuda}
\affiliation{Neutron Scattering Division, Oak Ridge National Laboratory, Oak Ridge, Tennessee 37831, USA}
\author{Yang Zhao}
\affiliation{NIST Center for Neutron Research, National Institute of Standards and Technology, Gaithersburg, Maryland 20899, USA}
\affiliation{Department of Materials Science and Engineering, University of Maryland, College Park, Maryland 20742, USA}
\author{J. W. Lynn}
\affiliation{NIST Center for Neutron Research, National Institute of Standards and Technology, Gaithersburg, Maryland 20899, USA}
\author{Zhiping Yin}
\affiliation{Department of Physics and Center for Advanced Quantum Studies, Beijing Normal University, Beijing 100875, China}
\author{Jun Zhao$^\ast$}
\affiliation{State Key Laboratory of Surface Physics and Department of Physics, Fudan University, Shanghai 200433, China}
\affiliation{Collaborative Innovation Center of Advanced Microstructures, Nanjing, 210093, China}
\begin{abstract}
{ We report neutron scattering measurements of single-crystalline YFe$_2$Ge$_2$ in the normal state, which has the same crystal structure to the 122 family of iron pnictide superconductors. YFe$_2$Ge$_2$ does not exhibit long range magnetic order, but exhibits strong spin fluctuations. Like the iron pnictides, YFe$_2$Ge$_2$ displays anisotropic stripe-type antiferromagnetic spin fluctuations at ($\pi$, $0$, $\pi$). More interesting, however, is the observation of strong spin fluctuations at the \textit{in-plane ferromagnetic} wavevector ($0$, $0$, $\pi$). These ferromagnetic spin fluctuations are isotropic in the ($H$, $K$) plane, whose intensity exceeds that of stripe spin fluctuations. Both the ferromagnetic and stripe spin fluctuations remain gapless down to the lowest measured energies. Our results naturally explain the absence of magnetic order in YFe$_2$Ge$_2$ and also imply that the ferromagnetic correlations may be a key ingredient for iron-based materials.}
\end{abstract}
\maketitle

Spin fluctuations can provide the pairing force for magnetic unconventional superconductors. In cuprate superconductors, it is widely believed that the pairing symmetry is ubiquitously $d$-wave, which is likely mediated by N\'{e}el spin fluctuations near ($\pi$, $\pi$) (square lattice unit cell) \cite{CuRev}. In iron-based superconductors containing both electron and hole Fermi surfaces, enormous previous efforts have been devoted to study the stripe spin fluctuations at ($\pi$, $0$), which tend to result in a sign-reversed s-wave pairing \cite{Dai2}. Interestingly, recent nuclear magnetic resonance (NMR) studies based on the analysis of the modified Korringa relation have suggested the coexistence of antiferromagnetic and ferromagnetic spin correlations in several iron-based superconductors \cite{FeNMR}, indicating that their magnetism could be more complicated than previously expected. However, ferromagnetic spin fluctuations associated with the FeAs/Se layer have thus far not been confirmed by inelastic neutron scattering experiments in any iron-based superconductor.\\
\indent
The recent discovery of superconductivity in the iron-germanide compound YFe$_2$Ge$_2$ ($T_c$ $\sim 1.8$ K), which does not belong to the iron pnictide or chalcogenide families, provides a new opportunity to investigate the nature of spin fluctuations and their possible link with superconductivity in iron-based materials \cite{Zou,Chen}. YFe$_2$Ge$_2$ has the same layered crystal structure to the 122 family of iron pnictides, namely the ThCr$_2$Si$_2$-type structure, but possesses a shorter inter-layer distance \cite{Caxis}. Nevertheless, angle resolved photoemission spectroscopy (ARPES) has revealed a quasi-two dimensional electronic structure with two hole pockets at the zone center and one electron pocket at the zone edge in YFe$_2$Ge$_2$ \cite{Xu}, which is similar to that of 122 iron pnictides. Density functional theory calculations predict an in-plane ferromagnetic ordered state in YFe$_2$Ge$_2$ \cite{Subedi,Singh}, resembling its sister compound LuFe$_2$Ge$_2$ \cite{LuFeGe}. However, no indication of any magnetic phase transition was observed by thermodynamic measurements down to the lowest measured temperature in YFe$_2$Ge$_2$ \cite{Zou,Chen,Canfield}. Instead, evidence for fluctuating magnetic moments in the core level photoemission spectroscopy measurements indicates the system is close to magnetic instabilities \cite{Sirica}. Moreover, NMR measurements suggested the coexistence of antiferromagnetic and ferromagnetic spin fluctuations in polycrystalline YFe$_2$(Ge,Si)$_2$ \cite{YFeGeSi}. These results together with the non-Fermi-liquid behavior of the resistivity/specific heat and the extremely disorder-sensitive superconductivity favor an unconventional pairing mechanism \cite{Zou,Chen,Canfield2}. Indeed, theoretical studies have proposed several unconventional pairing models, including a singlet pairing with an $s\pm$ gap function mediated through antiferromagnetic fluctuations \cite{Subedi}, and a triplet pairing associated with the in-plane ferromagnetic fluctuations \cite{Singh}. Thus, the elucidation of the nature of the spin fluctuations is pivotal to understanding the magnetism and superconductivity in this class of material.\\
\indent
We use neutron scattering to measure the spin fluctuations in YFe$_2$Ge$_2$ single crystals over the entire Brillouin zone. The experiments were carried out on the AMATERAS cold neutron disk chopper spectrometer at the Japan Proton Accelerator Research Complex \cite{AMATERAS}, ThALES three axis low energy spectrometer at the Institut Laue-Langevin, Grenoble, France, BT-7 triple-axis spectrometer at the NIST Center for Neutron Research \cite{jeff}, and HB-1 triple-axis spectrometer at high Flux Isotope Reactor, Oak Ridge National Laboratory. Our YFe$_2$Ge$_2$ single crystals were synthesized using the Sn-flux method. The crystals have a plate-like shape and show a residual resistivity ratio (RRR) of $\sim$ 40-60. According to ref.~\onlinecite{Chen}, the samples with RRR between 20 and 70 were non-superconducting or partially superconducting; bulk superconductivity was only observed in samples with RRR exceeding 70. This suggests that the superconductivity is extremely sensitive to disorder \cite{Chen}. Similar behavior has also been observed in the putative triplet $p$-wave superconductor Sr$_2$RuO$_4$ \cite{Sr2RuO4}. Nevertheless, our powder X-ray diffraction measurements on ground YFe$_2$Ge$_2$ single crystals found a single phase with no detectable impurities; the Rietveld refined lattice parameters \cite{Supplementary} are consistent with those of bulk superconducting samples in ref.~\onlinecite{Chen}. We have co-aligned around 200 pieces of single crystals with a total mass of 4 grams to achieve high counting statistics for the inelastic neutron scattering experiments. Since our measurements were performed in the normal state, the variation in superconducting properties induced by disorder is not expected to affect the spin excitation spectrum significantly, which is commonly the case in magnetic superconductors such as Sr$_2$RuO$_4$ \cite{SRO1,SRO2}.

\begin{figure}[h]
\centering
\includegraphics[scale=0.35]{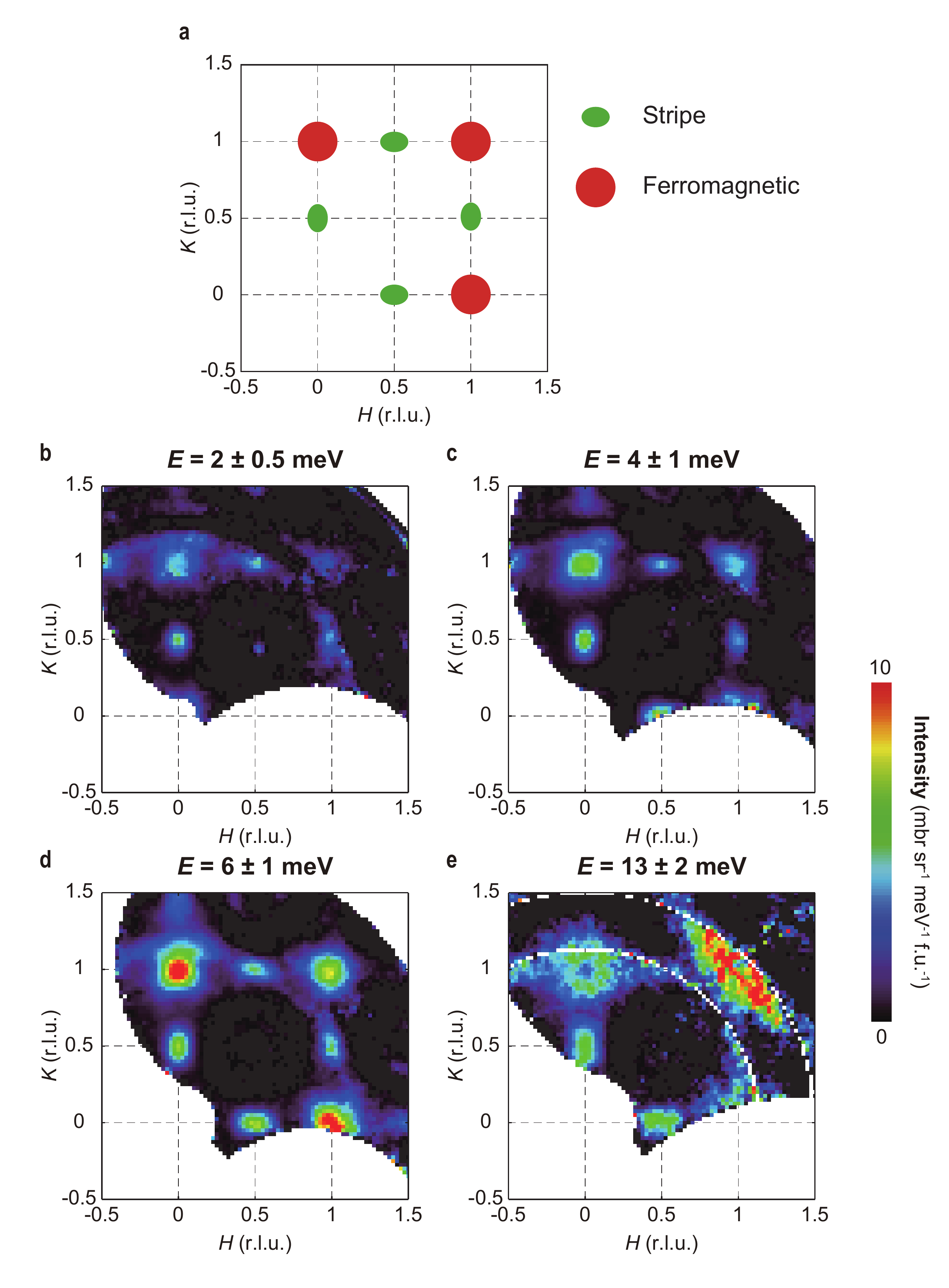}
\caption{Momentum dependence of the spin fluctuations in YFe$_2$Ge$_2$ at 4 K. (a) Schematic representation of the stripe and ferromagnetic spin fluctuations in the ($H$,$K$) plane. (b-e) Constant-energy images at indicated energies. The measurements (b)-(d) and (e) were carried out on AMATERAS using the incident neutron energies of 15 meV and 42 meV, respectively. The data were analyzed using the \textsc{Horace} program \cite{Horace}. The $|$\textbf{Q}$|$-dependent background has been subtracted using a similar method described in ref.~\onlinecite{Wang2}. The intensities are normalized to absolute units with acoustic phonons.}
\end{figure}

Fig. 1 shows a contour plot of the spin fluctuations in the (${H, K, 0.5}$) plane at 4 K. Clear scattering appears at \textbf{Q} = (0, 0.5, 0.5) and equivalent positions, corresponding to the stripe wavevector (1-Fe unit cell). The magnetic scattering is anisotropic in the (${H, K}$) plane and elongates along the longitudinal direction with respect to the reduced momentum transfer $\bf{q}$. This behavior resembles that of the hole-doped iron pnictides but differs from that of the electron-doped iron pnictides in which the scattering pattern is transversely elongated \cite{Dai2}. It should be noted that the nominal electron occupation of YFe$_2$Ge$_2$ for Fe is hole-doped 3\textit{d}$^{5.5}$, which seems consistent with the ARPES measurements that revealed two hole pockets at the zone center \cite{Xu}. However, it was also suggested that the relatively short Ge-Ge distance along the \textit{c} axis may lead to covalent bonds between Ge ions [Ge-Ge]$^{6-}$ \cite{Subedi}; the corresponding electron occupation of Fe therefore becomes electron-doped 3\textit{d}$^{6.5}$. It will be interesting to determine the exact electron concentration of YFe$_2$Ge$_2$ to see whether or how it can fit into the phase diagram of the carrier-doped 122 iron pnictides. More notably, in addition to the in-plane stripe-type antiferromagnetic spin fluctuations, much stronger spin responses are observed at \textbf{Q} = (0, 1, 0.5) and equivalent positions (Fig. 1), corresponding to the in-plane ferromagnetic wavevector. In contrast to the anisotropic stripe spin fluctuations, the ferromagnetic spin fluctuations are nearly isotropic along the $H$ and $K$ directions. With increasing energy, the ferromagnetic fluctuations disperse outwards, forming a nearly isotropic ring at 13 meV. The signals at both the stripe wavevector and ferromagnetic wavevector weaken with increasing $\left|\textbf{Q}\right|$ because of the reduced magnetic form factor, indicating their magnetic origin. On the other hand, no magnetic Bragg peaks at either the stripe or ferromagnetic wavevector were observed at temperatures down to 2 K (not shown), suggesting a non-magnetic ground state.

\begin{figure}[h]
\centering
\includegraphics[scale=0.3]{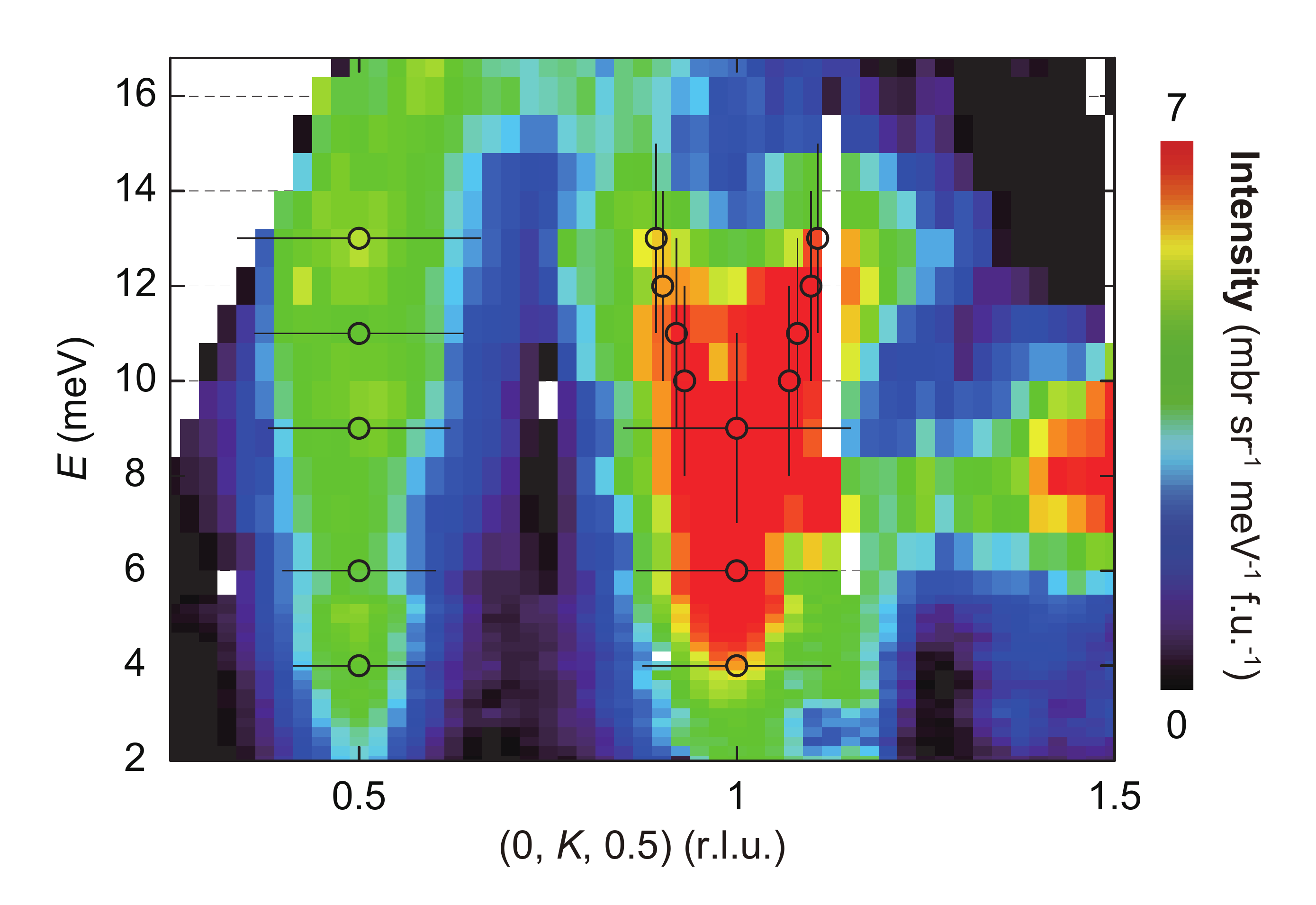}
\caption{Dispersions of the stripe and the ferromagnetic spin fluctuations in YFe$_2$Ge$_2$ at 4 K. The low energy (\textit{E}$<$5.6meV) and the high energy (5.6meV$<$\textit{E}$<$16.8meV) part were measured on AMATERAS with incident neutron energies of 15 meV and 42 meV, respectively. The $|$\textbf{Q}$|$-dependent background subtraction and intensities normalization were performed in the same way as in Fig. 1. The open circles represent for the peak positions determined from the Gaussian fitting of the constant-energy scans. The horizontal bars show the full-width at half maximum (FWHM) of the peaks, while the vertical bars indicate the range of integrated energies.}

\end{figure}

In order to elucidate the dispersion of the spin fluctuations, we present \textit{E}-\textbf{Q} relationships in Fig. 2. The stripe spin fluctuations exhibit a cone-like dispersion in the projection along the $K$ direction (or equivalently the $H$ directions for the (0.5, 0, 0.5) peak), consistent with the constant energy scans in Fig. 3. The scans along the longitudinal and transverse directions at the stripe wavevector are anisotropic (Fig. 3a-l), with a shorter dynamical spin correlation length along the longitudinal direction (Fig. 3g-l), ), versus steeper dispersion and longer correlation length along the transverse direction (Fig. 3a-f). This further confirms the anisotropy of the stripe spin fluctuation seen in the constant energy images in Fig. 1.

As for the ferromagnetic fluctuations, two branches of scattering arising from (0, 1, 0.5) can be clearly observed (Fig. 2). This can be illustrated more quantitatively in Fig. 3m-r. At $E = 2$ meV, a well-defined peak occurs centered at the ferromagnetic wavevector, which is broadened and then evolves into a pair of peaks with increasing energy. The peak positions and peak widths are determined by fitting the constant energy scans with Gaussian profiles (Fig. 3a-r), which are consistent with the contour plot in Fig. 2. We attempted but failed to fit the whole spectrum using a linear spin-wave theory for the Heisenberg model in either a pure stripe-type or a pure in-plane ferromagnetic order, because neither of them can account for the observed strong spin excitations at both wavevectors. The spin correlations along the $c$-axis are illustrated by constant energy scans along the $L$ direction in Fig. 3s,3t, where a clear $L$ modulation of the scattering intensity is observed. Such behavior is commonly seen in 122 iron pnictides where the inter-layer coupling is not negligible. Both the stripe and ferromagnetic spin fluctuation spectra exhibit a maximum intensity at $L$ = 0.5, 1.5, 2.5..., indicating an antiferromagnetic inter-layer coupling, which is consistent with the density functional theory calculations \cite{Subedi,Singh}.

\begin{figure}[h]
\centering
\includegraphics[scale=0.3]{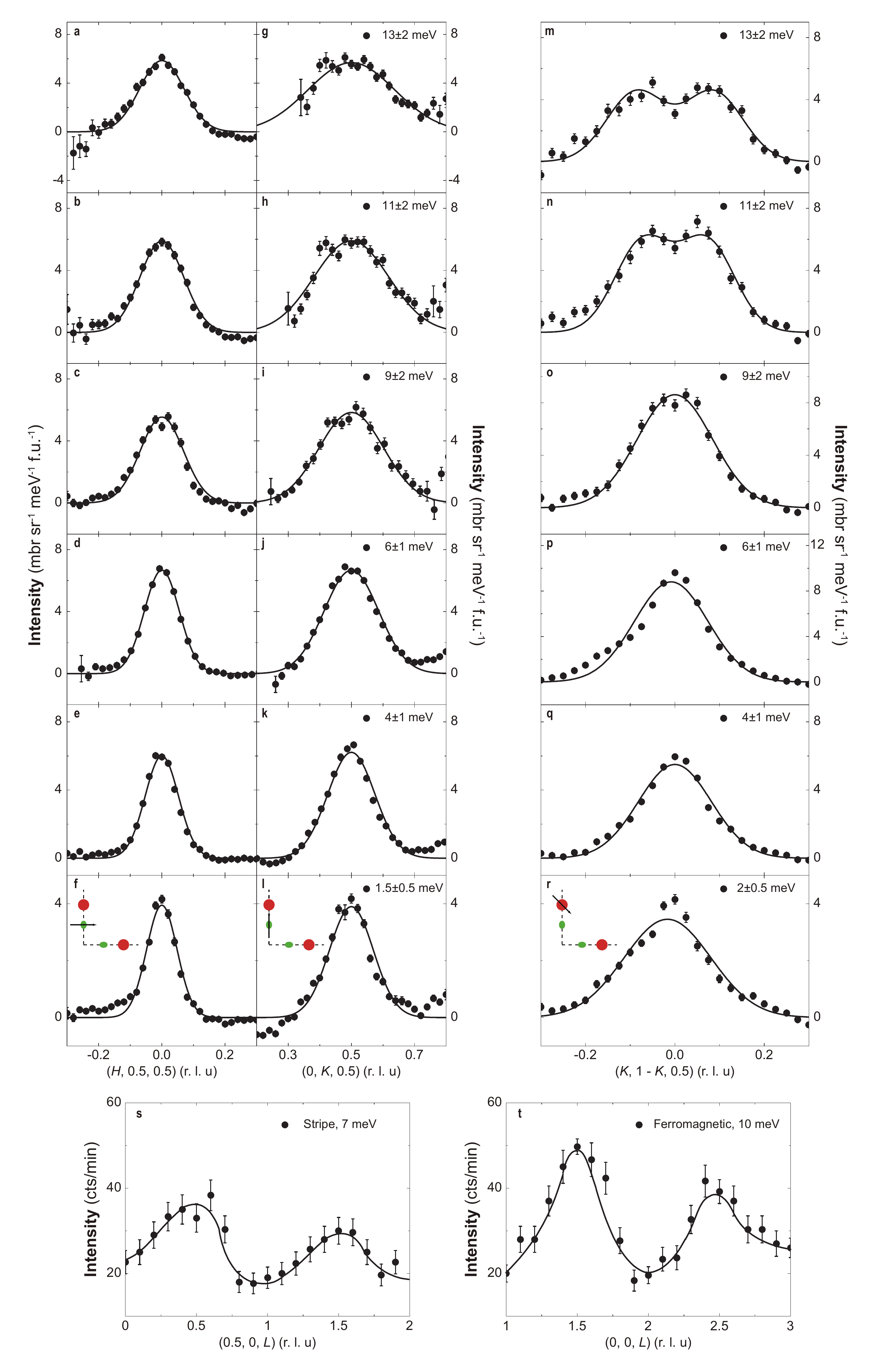}

\caption{Constant-energy scans of the stripe and ferromagnetic spin fluctuations in YFe$_2$Ge$_2$ at 4 K. (f, l), ((d)-(e), (j)-(k), (p)-(r)) and ((a)-(c), (g)-(i), (m)-(o)) were collected on AMATERAS with the incident neutron energies of 8meV, 15 meV and 42 meV, respectively. (s) and (t) were measured on HB-1 triple-axis spectrometer. (a)-(f), Background subtracted constant-energy scans through the stripe spin fluctuation along the \textit{H} (transverse) direction. (g)-(l), Background subtracted constant-energy scans for the stripe spin fluctuation along the \textit{K} (longitudinal) direction. (m)-(r), Background subtracted constant-energy scans for the in-plane ferromagnetic spin fluctuation along the (\textit{K}, \textit{1-K}) direction. (s)-(t), Constant-energy scans along the \textit{L} direction for the stripe and the in-plane ferromagnetic spin fluctuations.}

\end{figure}

By normalizing the spin excitation intensities with acoustic phonon modes \cite{Normalize}, we can calculate the momentum integrated local susceptibility $\chi''(\omega)$ in absolute units. Fig. 4a shows that the ferromagnetic spectral weight is much larger than that of the stripe spin excitations, implying that the system could be closer to ferromagnetic order than stripe order. Both the ferromagnetic and stripe spin fluctuations remain gapless down to the lowest measured energy (0.5 meV) (Fig. 4a). We note that the total local susceptibility in YFe$_2$Ge$_2$ is comparable to that of 122 iron pnictides at the energies measured \cite{Dai2}. This is not surprising since both systems have similar strength of electron correlations \cite{Xu}.

\begin{figure}[h]
\centering
\includegraphics[scale=0.2]{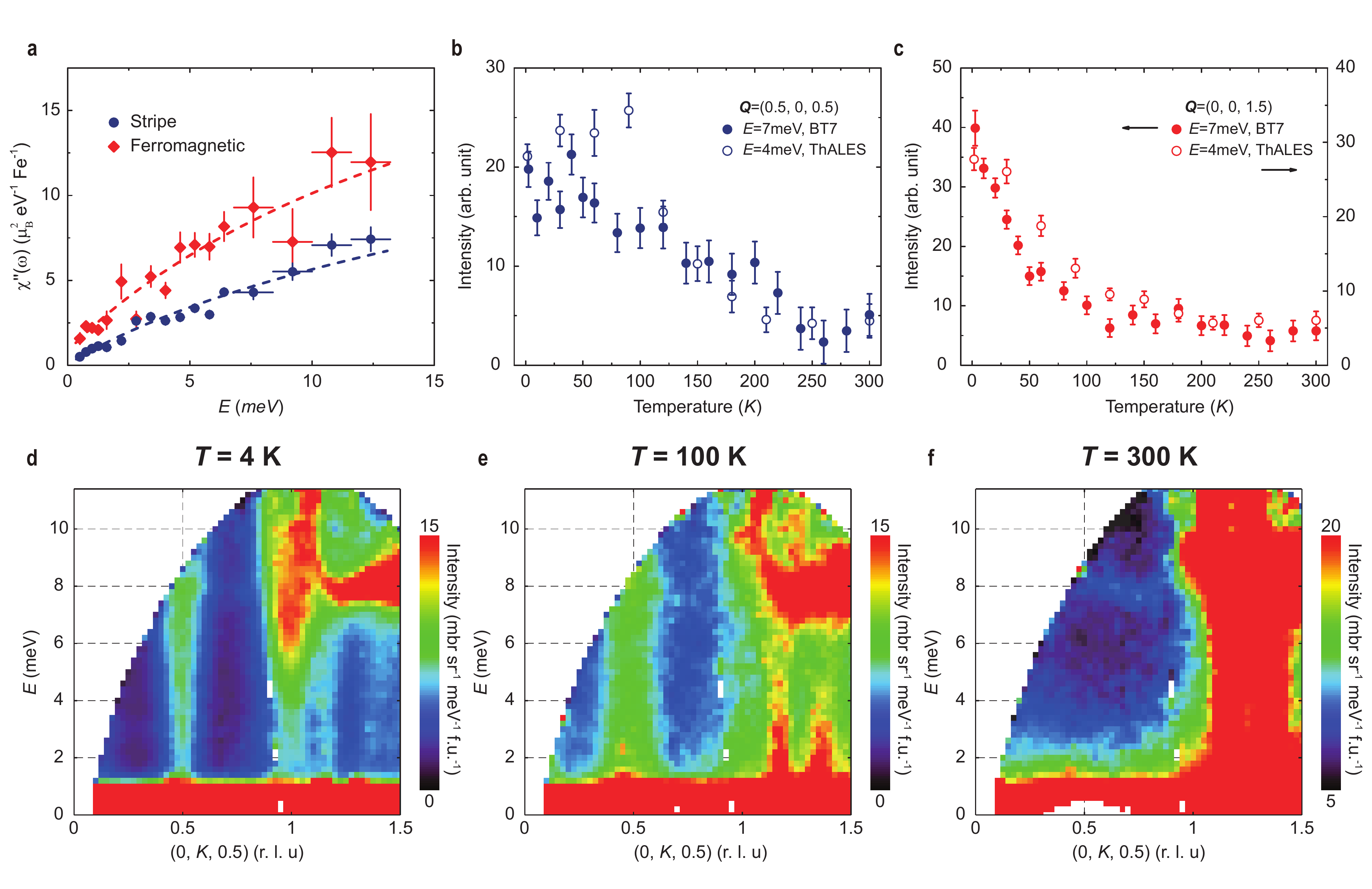}

\caption{Momentum integrated local susceptibility and temperature dependence of the stripe and ferromagnetic spin fluctuations in YFe$_2$Ge$_2$. (a) $\chi''(\omega)$ of the stripe and ferromagnetic spin fluctuations in YFe$_2$Ge$_2$ at 4 K. The absolute units were obtained via normalization for the acoustic phonon mode. (b)-(c) The \textit{E}=4 meV data (open circles) and the \textit{E}=7 meV (filled circles) was collected on ThALES and BT-7, respectively. The error bars indicate one standard deviation. (d)-(f) Raw data for the \textit{E}-\textbf{Q} relationship at different temperatures. The data were measured on AMATERAS with the incident neutron energy of 15 meV.}
\end{figure}
More insight into the nature of these two types of spin fluctuations can be obtained by measuring their temperature dependence. As is shown in Fig. 4b, the intensity of the stripe spin fluctuations shows a gradual decrease with increasing temperature from 4 K to 300 K. Such behavior is consistent with its magnetic origin and resembles the low energy response in other iron-based superconductors \cite{Wang}. On the other hand, warming from low temperature, the ferromagnetic spin fluctuations exhibit a steeper reduction with the correlations becoming negligible above $\sim 200$ K (Fig. 4b). The different temperature dependence of spin fluctuations could be due to the competition between these two magnetic instabilities; the ferromagnetic instability is favored at low temperature (4 K) while the stripe spin fluctuations is more stable against thermal fluctuations at relatively higher temperature ($>200$ K). This is further confirmed by the $E$-$\bf{Q}$ slices at various temperatures where ferromagnetic spin fluctuations disappear more rapidly with increasing temperature (Fig. 4d-f). On the other hand, the intensity of the phonon scattering at large $q$ increases significantly on warming because of the Bose population factor.

The unambiguous observation of stripe and ferromagnetic spin fluctuations within the Fe plane in one compound has important implications for understanding the nature of the magnetism and superconductivity in iron-based superconductors. Although density functional theory calculations and NMR measurements have suggested the coexistence of ferromagnetic and antiferromagnetic correlations in several classes of iron-based superconductors \cite{FeNMR,Subedi,Singh,YFeGeSi,Singh2,Mazin,NLWang,Mazin2}, YFe$_2$Ge$_2$ is the first compound in which ferromagnetic spin fluctuations are observed in inelastic neutron scattering experiments. Ferromagnetic fluctuations in general tend to mediate $p$-wave triplet pairing and introduce pair breaking for singlet superconductivity while antiferromagnetic fluctuations behave in an opposite manner \cite{Singh}. In this sense, the observation of stronger ferromagnetic spin fluctuations seems to favor a triplet pairing, while singlet pairing associated with stripe spin fluctuations could be a competitor. Regardless of the pairing symmetry, this may account for the relatively low $T_c$ in YFe$_2$Ge$_2$, since neither the ferromagnetic nor stripe spin fluctuation is completely dominating. To accurately determine the pairing symmetry, detailed microscopic measurements of the superconducting gap structure in the superconducting state are required.

The competition of the ferromagnetic and the stripe magnetic instabilities also provides a natural understanding of the absence of magnetic order in YFe$_2$Ge$_2$. This is in analogy to FeSe where the stripe spin fluctuations coexist with the N\'{e}el spin fluctuations \cite{Wang2}, resulting in a frustrated nematic ground state. However, in contrast to FeSe where nematic order develops at $\sim 90$ K, no evidence for nematicity has been revealed so far in YFe$_2$Ge$_2$. This could be due to the fact that ferromagnetic fluctuations are stronger in YFe$_2$Ge$_2$ while nematicity is generally coupled with stripe spin fluctuations. We note that electron doping or applying pressure in FeSe may partially remove the magnetic frustration and significantly enhance $T_c$ \cite{Wang2,Pressure,Xue,LiFeO,Doping2,Gating}. It therefore would be interesting to see if doping/pressure can do the same in YFe$_2$Ge$_2$.

In summary, we have presented a neutron scattering study of spin correlations of YFe$_2$Ge$_2$ single crystals. Unlike previous neutron studies focusing on antiferromagnetic spin fluctuations in iron pnictides/chalcgenides, we find the coexistence of strong stripe and ferromagnetic spin fluctuations within the Fe plane in YFe$_2$Ge$_2$, which suggests that it is a competition between the two that determines the fluctuating magnetic ground state and its relation to the pairing mechanism. By determining the momentum integrated magnetic spectral weight in absolute units, we show that the ferromagnetic spin fluctuations are stronger than stripe spin fluctuations in the low temperature regime, which possibly points to an unconventional triplet pairing. Our results, together with reported NMR data \cite{FeNMR,YFeGeSi}, imply that the competition between ferromagnetic and antiferromagnetic spin fluctuations could be a general property of iron-based superconductors, and call for a detailed search for the ferromagnetic correlations in other iron-based superconductors.

We thank J. T. Park for the assistance with the ThALES experiment, and D. F. Xu and D. L. Feng for discussions. This work was supported by the Innovation Program of Shanghai Municipal Education Commission (grant number 2017-01-07-00-07-E00018), the Ministry of Science and Technology of China (Program 973: 2015CB921302) and the National Key R\&D Program of the MOST of China (grant number 2016YFA0300203). A portion of this research used resources at the High Flux Isotope Reactor, a DOE Office of Science User Facility operated by the Oak Ridge National Laboratory. The experiment at AMATERAS was carried out under approval of J-PARC Center (Proposal No.2015A0126). H.W., Q.W., and Y.S. contributed equally to this work.

$\ast$ Correspondence and requests for materials should be addressed to J. Z. (zhaoj@fudan.edu.cn)

\end{document}